\documentclass[oribibl]{llncs}
\usepackage{multicol}
\usepackage{amsmath}
\usepackage{amssymb}
\usepackage{graphicx}
\usepackage{epsfig}
\usepackage{color}
\usepackage{hhline}
\usepackage{pdfsync}
\newcommand{\ignore}[1]{}

\newcommand{\boxtheorem}{\hfill $\Box$}
\newcommand{\nit}[1]{{\it #1}}

\usepackage{times}

\newcommand{\bul}{$\bullet$~}
\newcommand{\rdo}{\rho_{\!_D}\!}

\usepackage{graphicx}

\usepackage{algpseudocode}
\usepackage{algorithm}
\usepackage{epstopdf}
\usepackage{enumerate}
\usepackage{times}

\newcounter{lemmaA-counter}
\newcounter{propositionA-counter}
{\vskip \abovedisplayskip \refstepcounter{lemmaA-counter}%
\noindent {\bf Lemma A.\arabic{lemmaA-counter}.}}%

{\vskip \abovedisplayskip \refstepcounter{propositionA-counter}%
\noindent {\bf Proposition A.\arabic{propositionA-counter}.}}%

\newcommand{\mc}[1]{\mathcal{ #1}}

\newcommand{\bcq}{BCQ}
\newcommand{\cq}{CQ}

\newcommand{\mf}[1]{\mathfrak{ #1}}

\newcommand{\fo}{FO}

\newcommand{\red}[1]{{#1}}

\newcommand{\sfd}{{\sf d}}
\newcommand{\sfs}{{\sf s}}

\title{\vspace*{-1cm} {\bf The Causality/Repair Connection in Databases: Causality-Programs}\vspace{-3mm}}
\author{{\bf Leopoldo Bertossi}\thanks{Email: bertossi@scs.carleton.ca. \ Research supported by NSERC Discovery Grant \#06148.}}\ignore{\thanks{ Carleton Univ., \
School of Computer Science, Canada. \
bertossi@scs.carleton.ca.}}

\institute{ Carleton University, \
School of Computer Science, Ottawa, Canada. \
\vspace{-7mm}}

\begin{document}
\maketitle
\thispagestyle{empty}
\pagestyle{empty}

\vspace{1mm}

\ignore{

\vspace{-3mm}
\begin{figure}[ht]
\begin{center}
\vspace{-4mm}
 \includegraphics[width=9.5cm]{./fig/omdm} \vspace{-3mm}

 \caption{An \omd \ model with categorical relations, dimensional rules, and constraints}\label{fig:omdm} \vspace{-8mm}
\end{center}
\end{figure}
}

\begin{abstract}In this work, answer-set programs that specify repairs of databases are used as a basis for solving computational and reasoning problems about causes for query answers from databases. \vspace{-2mm}
\end{abstract}

\section{Introduction} \vspace{-2mm}

 Causality appears at the foundations of many scientific disciplines. In data and knowledge management, the need to represent and compute {\em causes} may be related to some
 form of
 {\em uncertainty} about the information at hand. More specifically
in data management, we need to understand
why certain  results, e.g. query answers, are obtained or not. Or why certain natural semantic conditions are not satisfied. These tasks become more prominent and difficult when dealing with large volumes of data.
One would expect the database to provide {\em explanations}, to understand, explore and make sense of the data, or to reconsider
queries and integrity constraints (ICs). Causes for data phenomena can be seen as a kind of explanations.

Seminal work on {\em causality in DBs}  introduced in \cite{Meliou2010a}, and building on work on causality as found in artificial   intelligence,  appeals to
the notions of counterfactuals, interventions and structural models \cite{Halpern05}. Actually, \cite{Meliou2010a} introduces the notions of: (a) a DB tuple as an {\em actual cause} for a query result, \ (b) a
{\em contingency set} for a cause, as a set of tuples that must accompany the cause for it to be such, and (c) the {\em responsibility} of a cause as a numerical measure of its strength  (building on \cite{Chockler04}).

Most of our research on causality in DBs   has been motivated by an attempt to understand causality from different angles of data and knowledge management. In \cite{tocs},
precise reductions between causality in DBs, DB repairs, and consistency-based diagnosis were established; and the relationships where investigated and exploited. In \cite{flairsExt}, causality in DBs was related to view-based DB updates and abductive diagnosis.
These are all interesting and fruitful connections among several forms of non-monotonic reasoning; each of them
reflecting  some form of  uncertainty about the information at hand. In the case of DB  repairs \cite{Bertossi2011}, it is about the uncertainty due the non-satisfaction of given ICs, which is represented by presence of possibly multiple intended repairs of the  inconsistent DB.

DB repairs can be specified by means of {\em answer-set programs} (or {\em disjunctive logic programs with stable model semantics}) \cite{gelfond}, the so-called {\em repair-programs}. Cf. \cite{monica,Bertossi2011} for repair-programs and additional references. In this work we exploit the reduction of DB causality to DB repairs established in \cite{tocs}, by taking advantage of repair programs for
specifying and computing causes, their contingency sets, and their responsibility degrees.
We show that that the resulting {\em causality-programs} have the necessary and sufficient expressive power to capture and compute not only causes, which can be done with less expressive programs \cite{Meliou2010a}, but specially minimal contingency sets and responsibilities (which can not). Causality programs can also be used for reasoning about causes.
Finally, we briefly show how causality-programs can be adapted to give an account of other forms of causality in DBs.

\vspace{-4mm}
\section{Background}

\vspace{-2mm}
\paragraph{Relational DBs.} \
  A relational schema $\mc{R}$ contains a domain, $\mc{C}$, of constants and a set, $\mc{P}$, of  predicates of finite arities. $\mc{R}$ gives rise to a language $\mf{L}(\mc{R})$ of first-order (FO)  predicate logic with built-in equality, $=$.  Variables are usually denoted by $x, y, z, ...$, and sequences thereof by $\bar{x}, ...$; and constants with $a, b, c, ...$, etc. An {\em atom} is of the form $P(t_1, \ldots, t_n)$, with $n$-ary $P \in \mc{P}$   and $t_1, \ldots, t_n$ {\em terms}, i.e. constants,  or variables.
  An atom is {\em ground} (aka. a tuple) if it contains no variables. A DB  instance, \red{$D$}, for $\mc{R}$ is a finite set of ground atoms; and it serves as an  \ignore{The {\em active domain} of a DB instance $D$, denoted ${\it Adom}(D)$, is the set of constants that appear in atoms of $D$.} interpretation structure for  $\mf{L}(\mc{R})$.

A {\em conjunctive query} (\cq) is a \fo \ formula,  $\mc{Q}(\bar{x})$, of the form \ $\exists  \bar{y}\;(P_1(\bar{x}_1)\wedge \dots \wedge P_m(\bar{x}_m))$,
 with $P_i \in \mc{P}$, and (distinct) free variables $\bar{x} := (\bigcup \bar{x}_i) \smallsetminus \bar{y}$. If $\mc{Q}$ has $n$ (free) variables,  $\bar{c} \in \mc{C}^n$ \ is an {\em answer} to $\mc{Q}$ from $D$ if $D \models \mc{Q}[\bar{c}]$, i.e.  $Q[\bar{c}]$ is true in $D$  when the variables in $\bar{x}$ are componentwise replaced by the values in $\bar{c}$. $\mc{Q}(D)$ denotes the set of answers to $\mc{Q}$ from $D$ $D$. $\mc{Q}$ is a {\em boolean conjunctive query} (\bcq) when $\bar{x}$ is empty; and when {\em true} in $D$,  $\mc{Q}(D) := \{\nit{true}\}$. Otherwise, it is {\em false}, and $\mc{Q}(D) := \emptyset$.

In this work we consider integrity constraints (ICs), i.e. sentences of $\mf{L}(\mc{R})$,  that are: (a) {\em denial constraints} \ (DCs), i.e.  of the form $\kappa\!:  \neg \exists \bar{x}(P_1(\bar{x}_1)\wedge \dots \wedge P_m(\bar{x}_m))$,
where $P_i \in \mc{P}$, and $\bar{x} = \bigcup \bar{x}_i$; and (b) {\em functional dependencies} \ (FDs), i.e. of the form  $\varphi\!:  \neg \exists \bar{x} (P(\bar{v},\bar{y}_1,z_1) \wedge P(\bar{v},\bar{y}_2,z_2) \wedge z_1 \neq z_2)$. Here,
$\bar{x} = \bar{y}_1 \cup \bar{y}_2 \cup \bar{v} \cup \{z_1, z_2\}$, and $z_1 \neq z_2$ is an abbreviation for $\neg z_1 = z_2$.\footnote{The variables in the atoms do not have to occur in the indicated order, but their positions should be in correspondence in the two atoms.} A {\em key constraint} \ (KC) is a conjunction of FDs: \  $\bigwedge_{j=1}^k \neg \exists \bar{x} (P(\bar{v},\bar{y}_1) \wedge P(\bar{v},\bar{y}_2) \wedge y_1^j \neq y_2^j)$,
with $k = |\bar{y_1}| = |\bar{y}_2|$.
\ A given schema may come with its set of ICs, and its instances are expected to satisfy them. If this is not the case, we say the instance is {\em inconsistent}.

\vspace{-2mm}
\paragraph{Causality in DBs.} \ A notion of {\em cause} as an explanation for a query result was introduced  in \ \ \cite{Meliou2010a}, as follows. For a
relational instance  ${D=D^n \cup D^x}$,  where ${D^n}$ and ${D^x}$ denote the mutually exclusive sets of { endogenous}  and { exogenous} tuples,
a tuple  ${\tau \in D^n}$ is called a
{\em counterfactual cause} for  a BCQ ${\mc{Q}}$,  if \ ${D\models \mc{Q}}$ \ and \ ${D\smallsetminus \{\tau\}  \not \models \mc{Q}}$. Now,
${\tau \in D^n}$ is an {\em actual cause} for  ${\mc{Q}}$
if there  exists ${\Gamma \subseteq D^n}$, called a {\em contingency set} for $\tau$,  such that ${\tau}$ is a  counterfactual cause for ${\mc{Q}}$ in ${D\smallsetminus \Gamma}$.
This definition is based on \cite{Halpern05}. % We denote with $ {\mc{CS}(D,\mc{Q})}$ \ the set of actual causes for ${\mc{Q}}$ in ${D}$.

The notion of {\em responsibility} reflects the relative degree of causality of a tuple for
a query result \cite{Meliou2010a} (based on \ \cite{Chockler04}).
 The { responsibility} of an actual cause ${\tau}$ for ${\mc{Q}}$, is ${\rho(\tau) \ := \ \frac{1}{|\Gamma| + 1}}$, where
${|\Gamma|}$ is the
size of a smallest contingency set for ${\tau}$. If $\tau$ is not an actual cause, $\rho(\tau):= 0$.
 Tuples with { higher responsibility} are stronger explanations.

{\em In the following we will assume all the tuples in a DB instance are endogenous.} \ (Cf. \cite{tocs} for the general case.) The notion of cause as defined above can be applied to monotonic queries, i.e whose
sets of answers may only grow when the DB grows \cite{tocs}.\footnote{E.g. CQs, unions of CQs (UCQs), Datalog queries are monotonic.} {\em In this work we concentrate only on conjunctive queries, possibly with $\neq$}.

\vspace{-1mm}
\begin{example}  \label{ex:cause}  Consider the relational  DB $D = \{R(a_4,a_3), R(a_2,a_1), R(a_3,a_3),$ $ S(a_4),$ $ S(a_2), S(a_3)\}$,  and the query \
${\mc{Q}\!: \ \exists x \exists y ( S(x) \land R(x, y) \land S(y))}$. \ It holds, ${D \models \mc{Q}}$.

%\begin{multicols}{2}
%{\small
\ignore{\begin{tabular}{l|c|c|} \hline
$R$  & A & B \\\hline
 & $a_4$ & $a_3$\\
& $a_2$ & $a_1$\\
& $a_3$ & $a_3$\\
 \hhline{~--}
\end{tabular} \hspace*{0.5cm}\begin{tabular}{l|c|c|}\hline
$S$  & A  \\\hline
 & $a_4$ \\
& $a_2$ \\
& $a_3$ \\ \hhline{~-}
\end{tabular} %}
}

${S(a_3)}$ is a counterfactual cause for ${\mc{Q}}$:
if ${S(a_3)}$ is removed from ${D}$,
 ${\mc{Q}}$ is no longer true. Its
%\end{multicols}
responsibility is ${1}$. So, it is an actual cause with empty contingency set.  ${R(a_4,a_3)}$ is an actual cause for ${\mc{Q}}$ with contingency set
${\{ R(a_3,a_3)\}}$:
 if ${R(a_3,a_3)}$ is removed from ${D}$, ${\mc{Q}}$ is still true, but further removing ${R(a_4,a_3)}$ makes ${\mc{Q}}$ false.
 The responsibility of ${R(a_4,a_3)}$ is ${\frac{1}{2}}$.
 ${R(a_3,a_3)}$ and ${S(a_4)}$ are actual causes, with responsibility  ${\frac{1}{2}}$. \boxtheorem
\end{example}

\vspace{-5mm}
\paragraph{Database repairs.} \ Cf. \cite{Bertossi2011} for a survey on DB repairs and consistent query answering in DBs.   We introduce the main ideas by means of an example. The ICs we consider in this work can be enforced only by deleting tuples from the DB (as opposed to inserting tuples). Repairing the DB by changing attribute values is also possible \cite{Bertossi2011,tkde,tplp}, \cite[sec. 7.4]{tocs}, but until further notice we will not consider this kind of repairs.

%\centerline{\epsfig{file = picReps.eps,width=8cm}}
\vspace{-1mm}
\begin{example} \label{ex:rep} \  The DB $D = \{P(a), P(e), Q(a,b), R(a,c)\}$ is inconsistent with respect to the (set of) {\em denial constraints} (DCs) \ $\kappa_1\!: \ \neg \exists x \exists y (P(x) \wedge Q(x,y))$, and \
$\kappa_2\!: \ \neg \exists x \exists y (P(x) \wedge R(x,y))$. It holds \  $D \not \models \{\kappa_1, \kappa_2\}$.

\ignore{
\begin{multicols}{2}
\hspace*{-0.5cm}{\small
\begin{tabular}{c|c|}\hline
$P$&A\\ \hline
&a\\
&e\\ \hhline{~-}
\end{tabular}~~
\begin{tabular}{c|c|c|}\hline
$Q$&A&B\\ \hline
& a & b\\ \hhline{~--}
\end{tabular}~~
\begin{tabular}{c|c|c|}\hline
$R$&A&C\\ \hline
& a & c\\ \hhline{~--}
\end{tabular} }
{\begin{eqnarray*}
\psi_1\!: \ \neg \exists x \exists y (P(x) \wedge Q(x,y)),\\
\psi_2\!: \ \neg \exists x \exists y (P(x) \wedge R(x,y)).
\end{eqnarray*}}
\end{multicols}  }
A {\em subset-repair},  in short an {\em S-repair}, of $D$ wrt. the set of DCs is a $\subseteq$-maximal subset of $D$ that is consistent, i.e.  no proper superset is consistent. The following are
S-repairs: ${D_1 = \{P(e), Q(a,b), R(a,b)\}}$ and ${D_2 = \{P(e), P(a)\}}$.
\ A {\em cardinality-repair},  in short a {\em C-repair}, of $D$ wrt. the set of DCs is a maximum-cardinality, consistent subset of $D$, i.e. no  subset of $D$ with larger cardinality is consistent.  $D_1$  is
the only C-repair. \boxtheorem
\end{example}

\vspace{-2mm}
For an instance $D$ and a set $\Sigma$ of DCs, the sets of S-repairs and C-repairs are denoted with $\nit{Srep}(D,\Sigma)$ and $\nit{Crep}(D,\Sigma)$, resp.

%\paragraph{Answer set programs.} \ (the stuff in the KR paper)

\vspace{-3mm}

\section{Causality Answer Set Programs}

%\subsection{Causes from repairs}

\vspace{-2mm}
\paragraph{Causes from repairs.} \ In \cite{tocs} it was shown that causes for queries can be obtained from DB repairs.
Consider the BCQ \ ${\mc{Q}\!: \exists \bar{x}(P_1(\bar{x}_1) \wedge \cdots \wedge P_m(\bar{x}_m))}$ that is (possibly unexpectedly) true in  $D$: \ $D \models \mc{Q}$. Actual causes for $\mc{Q}$, their  contingency sets, and responsibilities can be obtained from DB repairs. First,
$\neg \mc{Q}$ is logically equivalent to  the  DC: \vspace{-2mm}
\begin{equation}
{{\kappa(\mc{Q})}\!: \ \neg \exists \bar{x}(P_1(\bar{x}_1) \wedge \cdots \wedge P_m(\bar{x}_m))}. \label{eq:qkappa} \vspace{-1mm}
\end{equation}
So, if $\mc{Q}$ is true in $D$, \ $D$ is inconsistent wrt. $\kappa(\mc{Q})$, giving rise to repairs of $D$ wrt. $\kappa(\mc{Q})$.

Next, we build differences, containing a tuple $\tau$, between $D$ and  S-  or  C-repairs: \vspace{-2mm} \begin{eqnarray}
 \mbox{(a) } \ \nit{Dif}^s(D,\kappa(\mc{Q}), \tau) \ &=& \ \{ D \smallsetminus D'~|~ D' \in \nit{Srep}(D,\kappa(\mc{Q})), \  \tau \in (D\smallsetminus D')\}, \label{eq:s}\\
 \mbox{(b) } \ \nit{Dif}^c(D,\kappa(\mc{Q}), \tau) \ &=& \ \{ D \smallsetminus D'~|~ D' \in \nit{Crep}(D,\kappa(\mc{Q})), \ \tau \in (D\smallsetminus D')\}. \label{eq:c}
 \end{eqnarray}

\vspace{-1mm}
It holds \cite{tocs}: \ $\tau \in D$ is an {actual cause} for $\mc{Q}$ iff
$\nit{Dif}^s(D, \kappa(\mc{Q}), \tau) \not = \emptyset$. \ Furthermore, each S-repair $D'$ for which $(D\smallsetminus D') \in \nit{Dif}^s(D, \kappa(\mc{Q}), \tau)$ gives us $(D\smallsetminus (D' \cup \{\tau\}))$ as a subset-minimal contingency set for $\tau$. \ Also, if { $\nit{Dif}^s(D$  $\kappa(\mc{Q}),  \tau) = \emptyset$}, then {$\rho(\tau)=0$}.
 \ Otherwise, { $\rho(\tau)=\frac{1}{|s|}$}, where {  $s \in \nit{Dif}^s(D,$ $\kappa(\mc{Q}), \tau)$} and there is no { $s' \in \nit{Dif}^s(D,\kappa(\mc{Q}), \tau)$} with { $|s'| < |s|$}.
\ As a consequence we obtain that $\tau$ is a most responsible actual cause  for $\mc{Q}$ \ iff \
$\nit{Dif}^c\!(D,\kappa(\mc{Q}), \tau) \not = \emptyset$.

\vspace{-2mm}
\begin{example} (ex. \ref{ex:cause} cont.) \label{ex:kappa} \  With the same instance $D$ and query $\mc{Q}$, we consider the
DC \ $\kappa(\mc{Q})$:  \ $\neg \exists x\exists y( S(x)\wedge R(x, y)\wedge S(y))$, which is not satisfied by $D$.
\ Here, ${\nit{Srep}(D, \kappa(\mc{Q})) =\{D_1, D_2,D_3\}}$ and ${\nit{Crep}(D, \kappa(\mc{Q}))=\{D_1\}}$, with
$D_1=$ $ \{R(a_4,a_3),$ $ R(a_2,a_1), R(a_3,a_3), S(a_4), S(a_2)\}$, \  $D_2 = \{ R(a_2,a_1), S(a_4),S(a_2),$  $S(a_3)\}$, \  $D_3 =$ $\{R(a_4,a_3), R(a_2,a_1), S(a_2),S(a_3)\}$.

For tuple \ ${R(a_4,a_3)}$,  \ ${\nit{Dif}^s(D, \kappa(\mc{Q}), {R(a_4,a_3)})=\{D \smallsetminus D_2\}}$ $= \{ \{ R(a_4,a_3),$ \linebreak $ R(a_3,a_3)\} \}$. So,
 $R(a_4,a_3)$ is an actual cause,  with responsibility $\frac{1}{2}$. \ Similarly, $R(a_3,a_3)$ is an actual cause, with responsibility $\frac{1}{2}$.
\ For tuple ${S(a_3)}$,  \  $\nit{Dif}^c(D, \kappa(\mc{Q}), S(a_3)) =$ $ \{D \smallsetminus D_1\} =\{ S(a_3) \}$.
So, $S(a_3)$
is an actual cause,  with responsibility 1, i.e. a  {most responsible cause}. \boxtheorem
\end{example}

\vspace{-2mm}
It is also possible, the other way around, to characterize repairs in terms of causes and their contingency sets. Actually this connection can be used to obtain complexity results for
causality problems from repair-related computational problems \cite{tocs}. Most computational problems related to repairs, specially C-repairs, which are related to most responsible causes, are provably hard.
This is reflected in a high complexity for responsibility \cite{tocs} \ (see below for some more details).

\vspace{-3mm}
\paragraph{Answer-set programs for repairs.} \ Given a DB $D$ and a set of ICs, $\Sigma$, it is possible to specify the repairs of $D$ wrt. $\Sigma$ by means of an answer-set program (ASP)  $\Pi(D,\Sigma)$, in the sense that the set, $\nit{Mod}(\Pi(D,\Sigma))$, of its stable models is  in one-to-one correspondence with  $\nit{Srep}(D,\Sigma)$ \cite{monica,barcelo} (cf. \cite{Bertossi2011} for more references). In the following we
consider a single denial constraint $\kappa\!:  \neg \exists \bar{x}(P_1(\bar{x}_1)\wedge \dots \wedge P_m(\bar{x}_m))$.\footnote{It is possible to consider a combination of several DCs and FDs, corresponding to UCQs (possibly with $\neq$),  on the causality side \cite{tocs}.}

Although not necessary for repair purposes, it may be useful on the causality side having global unique tuple identifiers (tids), i.e. every tuple $R(\bar{c})$ in $D$ is represented as $R(t,\bar{c})$ for some integer $t$ that is not used by any other tuple in $D$. For the repair program we introduce a nickname predicate $R'$ for every predicate $R \in \mc{R}$ that has an extra, final attribute to hold an annotation from the set $\{\sf{d}, \sf{s}\}$, for ``delete" and ``stays", resp. \ Nickname predicates are used to represent and compute repairs.

The {\em repair-ASP}, $\Pi(D,\kappa)$, for $D$ and $\kappa$ contains all the tuples in $D$ as facts (with tids), plus the following rules: \vspace{-2mm}
\begin{eqnarray*}
P_1'(t_1,\bar{x}_1,\sfd)\vee \cdots \vee P_m'(t_n,\bar{x}_m,\sfd) &\leftarrow& P_1(t_1,\bar{x}_1), \dots, P_m(t_m,\bar{x}_m),\\
P_i'(t_i,\bar{x}_i,\sfs) &\leftarrow& P_i(t_i,\bar{x}_i), \ \nit{not} \ P_i'(t_i,\bar{x}_i,\sfd), \ i=1,\cdots,m.
\end{eqnarray*}

\vspace{-2mm}A stable   model $M$ of the program determines a repair $D'$ of $D$: \ $D' := \{P(\bar{c})~|$ \linebreak $P'(t,\bar{c},\sfs) \in M\}$, and every repair can be obtained in this way \cite{monica}.
For an FD, say $\varphi\!: \ \neg \exists xyz_1z_2vw(R(x,y,z_1,v) \wedge R(x,y,z_2,w) \wedge z_1 \neq z_2)$, which makes the third attribute functionally depend upon the first two, the repair program contains the rules: \vspace{-2mm}
{\small \begin{eqnarray*}
R'(t_1,x,y,z_1,v,\sfd) \vee R'(t_2,x,y,z_2,w,\sfd) &\leftarrow& R(t_1,x,y,z_1,v), R(t_2,x,y,z_2,w), z_1 \neq z_2.\\
R'(t,x,y,z,v,\sfs) &\leftarrow& R(t,x,y,z,v), \ \nit{not} \ R'(t,x,y,z,v,\sfd).
\end{eqnarray*} }

\vspace{-6mm}
 For DCs and FDs, the repair program can be made non-disjunctive by moving all the disjuncts but one, in turns, in negated form to the body of the rule \cite{monica,barcelo}. For example, the rule
$P(a) \vee R(b) \leftarrow \nit{Body}$, can be written as the two rules \ $P(a)  \leftarrow \nit{Body}, \nit{not} R(b)$ and $R(b) \leftarrow \nit{Body}, \nit{not} P(a)$. Still the resulting program can be {\em non-stratified}
if there is recursion via negation \cite{gelfond}, as in the case of FDs and  DCs with self-joins.

\vspace{-1mm}
\begin{example} (ex. \ref{ex:kappa} cont.) \label{ex:kappa2} \ For the DC \ $\kappa(\mc{Q})$:  \ $\neg \exists x\exists y( S(x)\wedge R(x, y)\wedge S(y))$, the repair-ASP contains the facts (with tids)
$R(1,a_4,a_3), R(2,a_2,a_1), R(3,a_3,a_3),$ $ S(4,a_4),$ \linebreak $S(5,a_2), S(6,a_3)$, and the rules: \vspace{-1mm}
{\small \begin{eqnarray*}
S'(t_1,x,\sfd) \vee R'(t_2,x,y,\sfd) \vee S'(t_3,y,\sfd) &\leftarrow& S(t_1,x), R(t_2,x, y),S(t_3,y),\\
S'(t,x,\sfs) &\leftarrow& S(t,x), \ \nit{not} \ S'(t,x,\sfd). \ \ \ \ \mbox{ etc. }
\end{eqnarray*}}

\vspace{-6mm}Repair $D_1$ is represented by the stable model $M_1$ containing  $R'(1,a_4,a_3,\sfs),$\linebreak  $R'(2,a_2,a_1,\sfs), R'(3,a_3,a_3,\sfs), S'(4,a_4,\sfs), S'(5,a_2,\sfs)$, and
$S'(6,a_3,\sfd)$. \boxtheorem
\end{example}

\vspace{-5mm}
\paragraph{Specifying causes with repair-ASPs.} \
According to  (\ref{eq:s}), we concentrate on the differences between the $D$ and its repairs, now represented by
$\{P(\bar{c})~|~P(t,\bar{c},\sfd) \in M\}$, for $M$ a stable model of the repair-program. They are used to compute actual causes and their $\subseteq$-minimal contingency sets, both identified by  tids. So, given the repair-ASP for a DC $\kappa(\mc{Q})$, a binary predicate $\nit{Cause}(\cdot,\cdot)$ will contain a tid for cause in its first argument, and a tid for a tuple belonging to its contingency set. Intuitively, $\nit{Cause}(t,t')$ says that $t$ is an actual cause, and $t'$ accompanies  $t$ as a member of the former's contingency set (as captured by the repair at hand or, equivalently, by the corresponding stable model).
More precisely, for each pair of predicates $P_i, P_j$ in the DC $\kappa(\mc{Q})$ as in (\ref{eq:qkappa}) (they could be the same if it has self-joins), introduce the rule \ $\nit{Cause}(t,t') \leftarrow P_i'(t,\bar{x}_i,\sfd),  P_j'(t',\bar{x}_j,\sfd), t\neq t'$, with the inequality condition only when $P_i$ and $P_j$ are the same.

\vspace{-2mm}
\begin{example} (ex. \ref{ex:kappa} and \ref{ex:kappa2} cont.) \ The causes for the query, represented by their tids, can be obtained by posing simple queries to the program under the {\em uncertain or brave} semantics that makes true what is true in {\em some} model of the repair-ASP.\footnote{As opposed to the {\em skeptical or cautious} semantics that sanctions as true what is true in {\em all} models. Both semantics as supported by the DLV system \cite{dlv}, to which we refer below.} In this case, $\Pi(D,\kappa(\mc{Q})) \models_\nit{brave} \nit{Ans}(t)$, where the auxiliary predicate is defined on top of $\Pi(D,$ $\kappa(\mc{Q}))$ by the rules: \ $\nit{Ans}(t) \leftarrow R'(t,x,y,\sfd)$ and $\nit{Ans}(t) \leftarrow S'(t,x,\sfd)$.

The repair-ASP  can be extended with the following rules to compute causes with contingency sets:
\vspace{-6mm}

\begin{multicols}{2}
{\footnotesize
\noindent $\nit{Cause}(t,t') \leftarrow S'(t,x,\sfd), R'(t',u,v,\sfd)$,\\
$\nit{Cause}(t,t') \leftarrow S'(t,x,\sfd), S'(t',u,\sfd), t\!\neq\! t'$,\\
$\nit{Cause}(t,t') \leftarrow R'(t,x,y,\sfd), S'(t',u,\sfd)$.%, \\
%\nit{Cause}(t,t') &\leftarrow& R'(t,x,y,\sfd), R'(t',u,v,\sfd), t\neq t'.
}

\hspace*{-7mm}\begin{minipage}[t]{0.5\textwidth}
\noindent
For the stable model $M_2$ corresponding to repair $D_2$, we obtain $\nit{Cause}(1,3)$ and $\nit{Cause}(3,1)$, from the repair difference $D \smallsetminus D_2 = \{R(a_4,a_3),R(a_3,a_3)\}$. \boxtheorem
\end{minipage}
\end{multicols}
\end{example}

\vspace{-2mm}We can use the DLV system \cite{dlv} to build the contingency set associated to a cause, by means of its extension, DLV-Complex \cite{calimeri08}, that supports set building, membership and union, as built-ins. For every atom $\nit{Cause}(t,t')$, we introduce the atom $\nit{Con}(t,$ $\{t'\})$, and the rule that computes the union of (partial) contingency sets as long as they differ by some element:

{\footnotesize  $\nit{Con}(T,\nit{\#union}(C_1,C_2)) \leftarrow \nit{Con}(T,C_1),\nit{Con}(T,C_2), \#\nit{member}(M,C_1),$}\\
\hspace*{9cm} {\footnotesize $ \nit{not} \ \#\nit{member}(M,C_2)$.}

The responsibility for an actual  cause $\tau$, with tid $t$, as associated to a given repair $D'$ (with $\tau \notin D'$), and then to a given model $M'$ of the extended repair-ASP, can be computed by counting the number of $t'$s for which
$\nit{Cause}(t,t') \in M'$. This responsibility will be maximum within a repair (or model): \
$\rho(t,M') := 1/(1+ |d(t,M')|)$, where $d(t,M') := \{\nit{Cause}(t,t') \in M'\}$. This value can be computed by means of the {\em count} function, supported by  DLV  \cite{aggreg},   as follows: \
$\mbox{\nit{pre-rho}}(T,N) \leftarrow  \#\nit{count}\{T' : \nit{Con}(T,T')\} = N$, followed by  the rule computing the responsibility: \ $\nit{rho}(T,M) \leftarrow M  * (\mbox{\nit{pre-rho}}(T,M) +1) = 1$. Or equivalently, via $1/|d(M)|$, with $d(M') := \{P(t',\bar{c},\sfd)~|~P(t',\bar{c},\sfd) \in M'\}$.

Each model $M$ of the program so far will return, for a given tuple (id) that is an actual cause, a {\em maximal-responsibility contingency set within that model}: no proper subset is a contingency set for the given cause. However, its cardinality may not correspond to the (global) {\em maximum}  responsibility for that tuple. For that we need to compute only maximum-cardinality repairs, i.e. C-repairs.

C-repairs can be specified by means of repair-ASPs
\cite{arenas} that contain {\em weak-program constraints} \cite{buca,aggreg}. In this case, we want repairs that minimize the number of deleted tuples. For each DB predicate $P$, we introduce the weak-constraint\footnote{Hard program-constraints, of the form \ $\leftarrow \nit{Body}$, eliminate the models where they are violated.}
\ $\Leftarrow  P(t,\bar{x}),  P'(t,\bar{x},\sfd)$. In a model $M$ the body can be satisfied, and then the program constraint violated, but the number of violations is kept to a minimum (among the models of the program without the weak-constraints). A repair-ASP with these weak constraints specifies repairs that minimize the number of deleted tuples; and {\em minimum-cardinality} contingency sets and maximum responsibilities can be computed, as above.

\vspace{-2mm}
\paragraph{Complexity.} \ Computing causes for CQs can be done in polynomial time in data \cite{Meliou2010a}, which was extended to UCQs in \cite{tocs}. As has been established in \cite{Meliou2010a,tocs}, the computational problems associated to contingency sets and responsibility are in the second level of the polynomial hierarchy (PH), in data complexity
\cite{dantsin}. On the other side, our causality-ASPs can be transformed into non-disjunctive, unstratified programs, whose reasoning tasks are also in the second level of the PH (in data). It is worth mentioning that the ASP approach to causality via repairs programs could be extended to deal with queries that are more complex than CQs or UCQs. (In \cite{tapp16} causality for queries that are conjunctions of literals was investigated; and in \cite{flairsExt} it was established that cause computation for Datalog queries can be in the second level of the PH.)

\vspace{-2mm}
\paragraph{Causality programs and ICs} \ The original causality setting in \cite{Meliou2010a} does not consider ICs. An extension of causality under ICs was proposed in \cite{flairsExt}. Under it,  the ICs have to be satisfied by the DBs involved, i.e. the initial one and those obtained by cause- and contingency-set deletions. When the query at hand is monotonic\footnote{I.e. the set of answers may only grow when the instance grows.}, monotonic ICs (e.g. denial constraints and FDs) are not much of an issue since they stay satisfied under deletions associated to causes. So, the most relevant ICs are non-monotonic, such as referential ICs, e.g. $\forall xy(R(x,y) \rightarrow S(x))$ in our running example. These ICs can be represented in a causality-program by means of (strong) program constraints. In the running example, we would have, for example, the
constraint: \ $\leftarrow R'(t,x,y,\sfs), \nit{not} \ S'(t',x,\sfs)$.\footnote{Or better, to make it {\em safe}, by a rule and a constraint: \ $\nit{aux}(x) \leftarrow S'(t',x,\sfs)$ and \ $\leftarrow R'(t,x,y,\sfs), \nit{not}
\ \nit{aux}(x)$.}

\vspace{-2mm}
\paragraph{Preferred causes and repairs.} \ In \cite{tocs}, generalized causes were introduced on the basis of arbitrary repair semantics (i.e. classes of preferred consistent subinstances, commonly under some maximality criterion), basically starting from the characterization in
(\ref{eq:s}) and (\ref{eq:c}), but using repairs of $D$ wrt. $\kappa(\mc{Q})$ in a class, $\nit{Rep}(D,\kappa(\mc{Q}))$, possibly different from $\nit{Srep}(D,\kappa(\mc{Q}))$ or   $\nit{Crep}(D,\kappa(\mc{Q}))$. As a particular case in \cite{tocs}, {\em causes based on changes of attribute values} (as opposed to tuple deletions) were defined. In that case, admissible updates are replacements of data values by null values, to break joins, in a minimal or minimum way. Those underlying DB repairs were used in \cite{tkde} to hide sensitive data that could be exposed through CQ answering; and corresponding repair programs were introduced. They could be used, as done earlier in this paper, as a basis to reason about- and compute the new resulting  causes (at the tuple or attribute-value level) and their contingency sets.\footnote{Cf. also \cite{tplp} for an alternative null-based repair semantics and its repair programs.}

\end{document}